\documentclass[preprint]{revtex4-1}
\usepackage{graphicx}
\usepackage{color}
\usepackage{natbib}
\graphicspath{{figures/}}
\newcommand{\fig}[4]{\begin{figure}
\centering
\includegraphics[width=#2\textwidth]{#1}
\caption{#3}
\label{#4}
\end{figure}}
\begin{document}
\title{Direct-write of free-form 3D nanostructures with controlled magnetic frustration}
\author{Lukas Keller$^1$}
\author{Mohanad K.\ I.\ Al Mamoori$^1$}
\author{Jonathan Pieper$^1$}
\author{Christian Gspan$^2$}
\author{Irina Stockem$^3$}
\author{Christian Schr\"oder$^4$}
\author{Sven Barth$^5$}
\author{Robert Winkler$^2$}
\author{Harald Plank$^6$}
\author{Merlin Pohlit$^1$}
\author{Jens M\"uller$^1$}
\author{Michael Huth$^1$}
\affiliation{$^1$Institute of Physics, Goethe University, Frankfurt am Main, Germany}
\affiliation{$^2$Graz Centre for Electron Microscopy, Graz, Austria}
\affiliation{$^3$Department of Physics, Chemistry, and Biology (IFM), Link\"oping University, Sweden}
\affiliation{$^4$Bielefeld Institute for Applied Materials Research, Bielefeld University of Applied Sciences, Bielefeld, Germany}
\affiliation{$^5$Vienna University of Technology, Institute of Materials Chemistry, Wien, Austria}
\affiliation{$^6$Institute for Electron Microscopy and Nanoanalysis, Graz University of Technology, Graz, Austria}
%
%
\begin{abstract}
Building nanotechnological analogues of naturally occurring magnetic structures has proven to be a powerful approach to studying topics like geometry-induced magnetic frustration and to provide model systems for statistical physics. Moreover, it practically allows to engineer novel physical properties by realizing artificial lattice geometries that are not accessible via natural crystallization or chemical synthesis. This has been accomplished with great success in the field of two-dimensional artificial spin ice systems with important branches reaching into the field of magnetic logic devices. Although first proposals have been made to advance into three dimensions (3D), established nanofabrication pathways based on electron beam lithography have not been adapted to obtain free-form 3D nanostructures. Here we demonstrate the direct-write fabrication of freestanding ferromagnetic 3D nano-architectures with full control over the degree of magnetic frustration. By employing micro-Hall sensing, we have determined the magnetic stray field generated by our free-form structures in an externally applied magnetic field and we have performed micromagnetic and macro-spin simulations to deduce the spatial magnetization profiles in the structures and analyze their switching behavior. Furthermore we show that the magnetic 3D elements can be combined with other 3D elements of different chemical composition and intrinsic material properties.
\end{abstract}
\maketitle
%
%
\section{Introduction}
Nanomagnetic structures are ubiquitous, as they form the basic functional elements in various applications, such as in magnetic storage and information processing, magnonics and spintronics, see e.\,g.\ \cite{Kruglyag2010_magnonics_review, Pulizzi2012_spintronics_nature_materials_insight, Demokritov2013_magnonics_book}. Nanomagnetic structures are traditionally planar, but recent work is expanding nanomagnetism into three dimensions and it has been generally recognized that in three-dimensional nanomagnets complex magnetic configurations with unprecedented properties become possible, see \cite{Pacheco2017_3d_nanomagnetism} for a recent review. In the narrower sense of magnetic information storage and processing, the advantages of extending the typically 2D structures into the third dimension for higher integration density have already been realized and have lead to developments such as the racetrack memory \cite{Parkin2008_racetrack_memory}. Further on, in so-called artificial spin ice systems \cite{Melko2001_dipolar_spin_ice, Wang2006_first_spinice, Ladak2010_artificial_spin_ice_monopole_defects, Mengotti2011_artificial_spin_ice_dirac_strings_kagome, Zhang2013_artificial_spin_ice_magnetic_charges, Nisoli2013_spinice_review, Gilbert2016_artificial_spin_ice_vertex_frustration, Drisko2017_artificial_spin_ice_topological_frustration}, that are currently subject of intensive research efforts, the actual limitation to lithographically defined 2D arrays of interacting ferromagnetic nano-islands prevents investigations of novel phases that can emerge from the more complex ground states of frustrated lattices in 3D. This is why first steps into 3D artificial spin ice systems are now being taken by combining multilayer techniques with sophisticated electron beam lithography (EBL) \cite{Chern2014_3D_spinice_proposal}. However, standard lithography techniques are intrinsically designed for 2D pattern formation and, consequently, they are barely suitable for the fabrication of free-form 3D nanostructures. Focused electron beam induced deposition (FEBID) follows a different approach to overcome this EBL-related limitation. It represents a highly flexible direct-write fabrication method which allows for excellent control in creating 3D structures very much like 3D printing on the nanometer scale. FEBID uses precursor gases which, being adsorbed on a surface, are dissociated in the focussed electron beam to form the deposit. In most cases the resulting structures are not simple phase-pure metals or oxides. Instead, the resulting nanostructures typically contain significant amounts of carbon, which is predominantly part of the precursor species. Also, FEBID is a highly complex process in which many parameters, such as electron beam energy and current, precursor flow and adsorption characteristics, precursor diffusion and beam steering strategy all influence the final deposit shape and the deposit's composition \cite{Dorp2008_febid_review, Huth2012_febid_review, Fowlkes2016_3D_FEBID}. However, intensive research over the last decade has pushed the capabilities of FEBID in two important areas. It is now possible to obtain fully metallic nanostructures of Fe, Co and FeCo-alloys \cite{DeTeresa2016_febid_magnetic_review} and also of Au and Pt \cite{Geier2014_Pt_purification_H2O, Sachser2014_PtC_purification_pulsed_O2, Villamor2015_Pt_purification_direct_O2, Shawrav2016_Au_with_H2O, Winkler2017_3D_plasmonic}. In addition, very recently the simulation-guided nano-manufacturing of 3D structures has matured to such a degree that even complex 3D objects can now be fabricated under controlled conditions \cite{Fowlkes2016_3D_FEBID}. For pillar-like Fe and Co structures this has recently been demonstrated in a detailed investigation of the sample composition \cite{Cordoba2016_3Dpillars}. The next important scientific development is now the synergistic combination of these two developments towards the realization of free-form magnetic 3D nanostructures \cite{Pacheco2013_Co_3D_simple, Navarro2017_Co_3D_pillars}. Here, we demonstrate this next step by showing different examples of free-form 3D magnetic nano-architectures with a focus on magnetic frustration effects. The structures have been directly written on a high-resolution micro-Hall sensor. We followed the magnetization switching behavior by measuring the associated magnetic stray field during external field sweeps using a sensor with dimensions adapted to the size of the magnetic structures. With the help of micromagnetic and macro-spin model simulations we are able to explain the observed complex switching behavior. 
%
%
\section{Results}
%
%
\subsection{Geometry and microstructure}
For the deposition of magnetic 3D nanostructures by FEBID (see Fig.\,\ref{fig_EBID_SEM_TEM}(a) for FEBID principle) we chose the recently introduced precursor HCo$_3$Fe(CO)$_{12}$, as this was shown in our previous work to yield deposits of high Co$_3$Fe metal content under beam conditions which are suitable for writing high-resolution structures, i.\,e.\ high beam voltage and low beam current (see methods section for details) \cite{Porrati2015_CoFe}. In view of very recent findings by Cordoba \textit{et al.} \cite{Cordoba2016_3Dpillars} it should be possible to obtain similar structures than the ones shown here with high metal content from the precursors Fe$_2$(CO)$_9$ and Co$_2$(CO)$_8$ under suitable beam conditions.

In Fig.\,\ref{fig_EBID_SEM_TEM} we show two scanning electron microscopy (SEM) images taken directly after the writing of $2\times 2$ arrays of Fe-Co nano-trees (b) and nano-cubes (c) onto the top Au gate of a GaAs/AlGaAs micro-Hall sensor (see methods section for details). These represent magnetic nanostructures with vertices that are connected to three and four neighboring vertices, respectively.
\fig{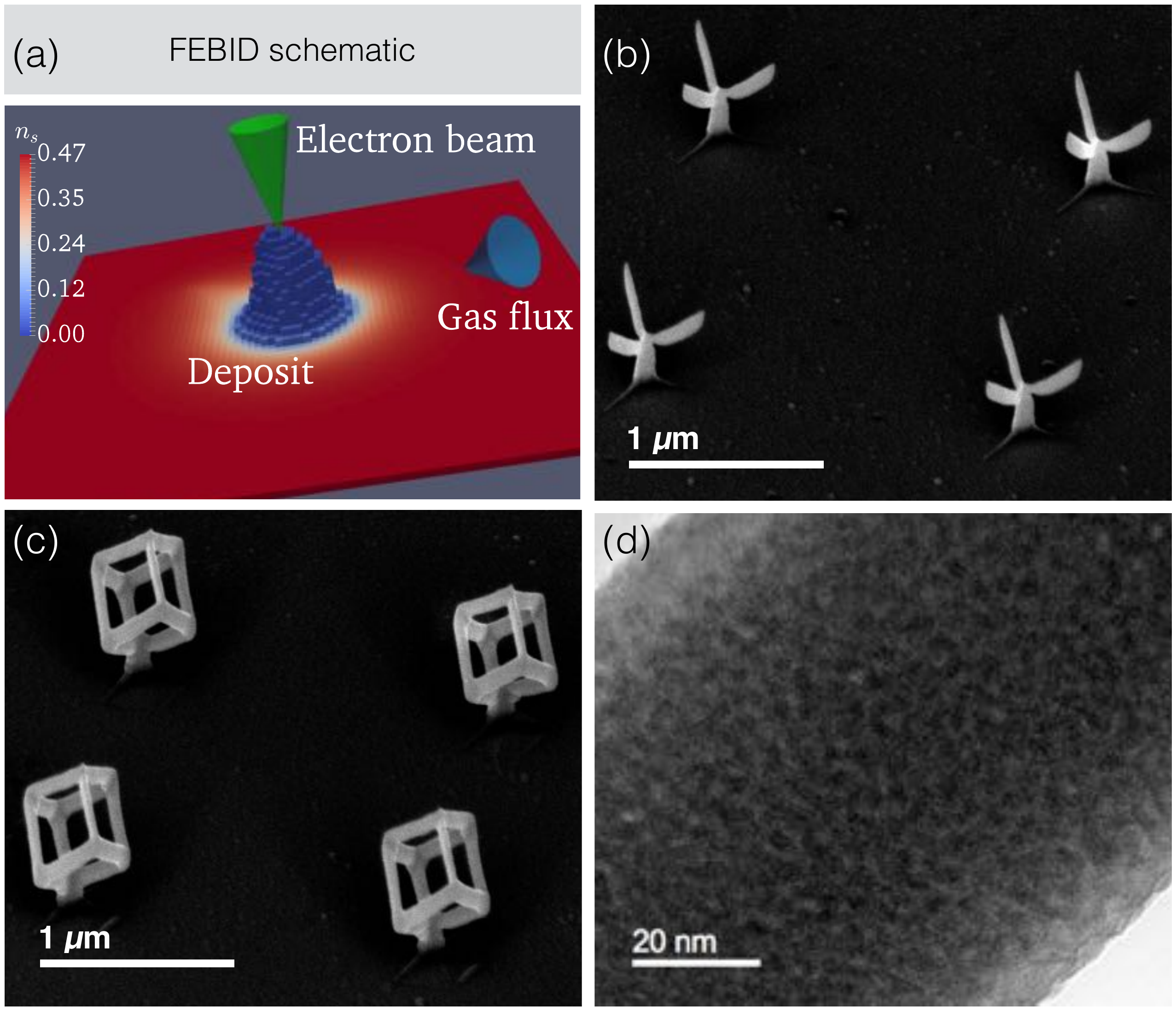}{0.95}{{\bf FEBID of 3D magnetic structures.} (a) Schematic of FEBID process. $n_s$ denotes the precursor coverage normalized to a maximum of one monolayer. See methods section for details. (b) and (c) SEM images of $2\times 2$ arrays of Fe-Co nano-trees (b) and nano-cubes (c).  (d) Transmission electron microscopy bright field image of nano-cube region demonstrating the nano-granular microstructure.}{fig_EBID_SEM_TEM}
In order to determine the microstructure and element distribution in the deposits we have performed transmission electron microscopy (TEM) experiments on nano-cubes grown onto metallic TEM grids. As the growth in 3D very sensitively depends on the precursor flux distribution \cite{Winkler2014_gas_flux_influence}, special care was taken to reproduce the nano-cube geometry as obtained on the micro-Hall sensor. Microstructure and element distribution contain the essential information for developing an appropriate micromagnetic simulation model, as described later. In Fig.\,\ref{fig_EBID_SEM_TEM}(d) we present a TEM bright field image of one of the nano-cube edges. The 3D edges reveal a homogeneously distributed nano-granular structure consisting of nano-crystallites of about $3\,$nm diameter on average. Additional chemical information is extracted from electron energy loss spectroscopy (EELS) and energy dispersive X-ray spectroscopy (EDXS). The latter reveal an overall composition of Co$_3$FeC$_{0.25}$O$_2$, corresponding to individual contents of $64\,$at$\%$ metal, $32\,$at$\%$ oxygen, and $4\,$at$\%$ carbon. Such a small carbon content can in fact be caused by unavoidable deposition of amorphous carbon during the EDXS analysis. This is a consequence of electron beam induced deposition of hydrocarbons adsorbed or chemisorbed on the sample surface and from the residual gas. During deposition the main carbon source is the precursor which contains a substantial amount of carbonyl groups. For planar deposits we found that the carbon to oxygen ratio is close to one \cite{Porrati2015_CoFe}. This indicates that carbonyl groups remain largely intact after dissociation and become part of the deposit, if their desorption is not sufficiently fast. Considering the small carbon content in the 3D structures in conjunction with the enhanced oxygen content, we argue that the oxygen is the result of a post-growth oxidation effect that occurred during sample transport and storage under ambient conditions before the TEM investigations were performed. Two further observations support this argument. First, scanning TEM EELS analysis of the oxygen content along the cross section of one of the nano-cube 3D edges reveals an enhanced oxygen content towards the surface (see \ref{fig_TEM_EELS}(b)).
\fig{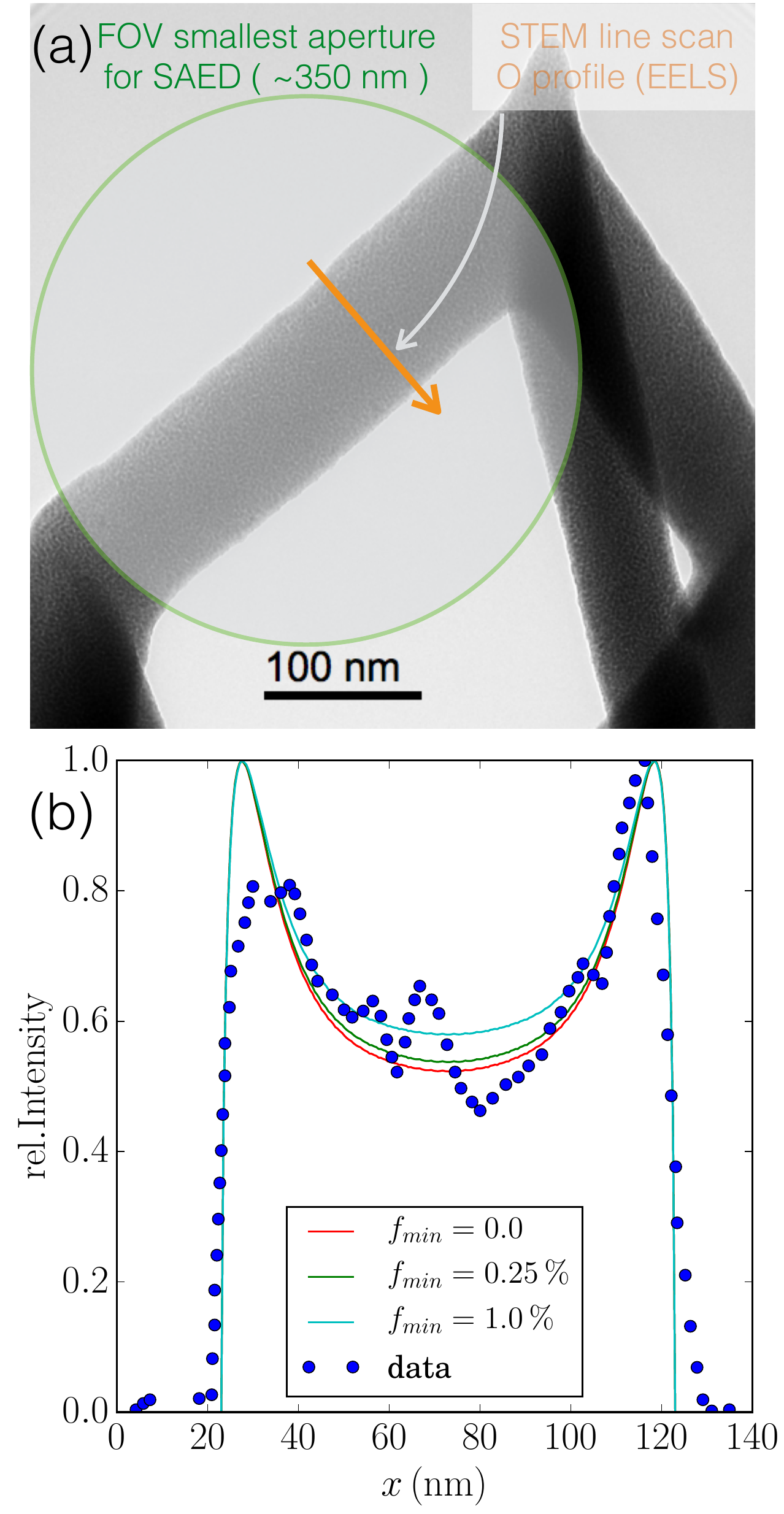}{0.5}{{\bf EELS signal oxygen distribution.} (a) TEM bright field image of an edge region of one nano-cube. The shaded area in the green circle indicates the region over which the selected area electron diffraction image, shown in Fig.\,\ref{fig_TEM_SAED} was taken (FOV: field of view). The orange arrow indicates the position and direction of the scanning TEM line along which EELS intensity data at the oxygen absorption energy were taken.  (b) EELS intensity vs.\ scanning TEM line as indicated in (a). The lines result from a fit of an exponentially decaying oxygen concentration from the surface to the center of the nano-cube arm, as detailed in the text. The fit parameters are: $R_1 = 50\,$nm (axis in scan-direction, i.\,e.\ $x$-direction),  $R_2 = 40\,$nm (axis in beam direction, i.\,e.\ $y$-direction), $\lambda = 5.0\,$nm, $f_{max} = 50\,$at$\%$.}{fig_TEM_EELS}
However, as EELS signal intensities do not directly provide quantitative concentration data we have to deconvolve the signal. We therefore start from the following relation for quantitative elemental analysis (see, e.\,g.\ \cite{Egerton2009_eels})
\begin{equation}
I(\beta, \Delta) = N I_0(\beta, \Delta) \sigma(\beta, \Delta)
\label{eq_eels_elemental_analysis_simple}
\end{equation}
for a given sample thickness, where $I$ denotes the EELS intensity of the element of interest with areal density $N$ in the energy range $\Delta$ beyond the element-specific threshold. $I_0$ is the integral of the low-loss spectrum up to $\Delta$, including the entire zero-loss peak, $\sigma(\beta, \Delta)$ is a partial cross section, and $\beta$ is the collection semiangle. The oxygen content may vary depending on the distance from the sample surface and the sample thickness will depend on the position of the EELS line scan. Tilt series via both, TEM and high-resolution SEM, reveal elongated instead of circular edge cross sections for 3D elements in agreement with previous studies \cite{Fowlkes2016_3D_FEBID, Winkler2017_3D_plasmonic}. To approach the real situation with a simplified analytical expression, we assume an elliptical cross-section of radii $R_1$ and $R_2$ for the two main axes of any of the nano-cube edges. From Eq.\,\ref{eq_eels_elemental_analysis_simple} we derive the following expression for the EELS intensity $I(\beta, \Delta; x)$ at any given position $x$ along the line scan shown in Fig.\,\ref{fig_TEM_EELS}(a) by integrating along the beam direction $+y$ in the limits $y_1(x)$ and $y_2(x)$ defined by the respective sample thickness at scan position $x$
\begin{equation}
\frac{I(\beta, \Delta; x)}{I_0(\beta, \Delta) \sigma(\beta, \Delta)} =  \int_{y_1(x)}^{y_2(x)} f(x, y) e^{-y/\Lambda} dy \quad\textnormal{with}\quad f(x, y) = f_{min} + \left( f_{max} - f_{min} \right) e^{-\xi(x, y)/\lambda}
\label{eq_oxygen_content_eels_fit_model}
\end{equation}
with $N(x, y) = f(x, y) dy$, $f(x, y)$ being the volume density or concentration of oxygen, and $\Lambda = 148\,$nm the inelastic mean free path. The form of the oxygen concentration function $f(x, y)$ is taken to be exponentially decaying from the surface towards the bulk, as Fig.\,\ref{fig_TEM_EELS}(b) indicates, so that the oxygen concentration decreases from the surface into the bulk. Correspondingly, $\xi(x, y)$ is the vertical distance of the point $(x, y)$ inside the cube edge from the surface.  The result of this fit, assuming different oxygen contents, is shown as solid lines in Fig.\,\ref{fig_TEM_EELS}(b). The measured line scan can be reproduced quite well by the fit and clearly indicates a very low oxygen content (below $1\,$at$\%$) in the center of the nano-cube edge that increases towards the surface to about $50\,$at$\%$ with a characteristic length of $\lambda = 5\,$nm. Given this, we assume a similar oxidation profile for the corresponding nano-cube and nano-tree structures used for the magnetic measurements.

A second, independent observation supports our assumption of a nearly $100\,$at$\%$ metal content in the bulk of our 3D nanostructures. Atom probe tomography on Fe- and Co-nano-pillars performed by Cordoba and collaborators showed that mass transport limited 3D growth provides favorable conditions for the complete desorption of carbonyl groups after dissociation \cite{Cordoba2016_3Dpillars}.

In order to elucidate the degree of crystallinity of the deposits we performed selective area electron diffraction (SAED). The smallest field of view accessible in our setup is indicated in \ref{fig_TEM_EELS}(c) by the shaded area (green circle). Consequently, we were not able to discriminate between the near-surface and bulk regions. This has to remain for future investigations. In Fig.\,\ref{fig_TEM_SAED} we show a diffraction image as measured and reference the diffraction rings with the corresponding scattering vectors. Apparently, the deposits are crystalline and we can attribute all diffraction rings to the $\alpha$-phase of the Co-Fe binary system (bcc) and a Co-rich spinel phase of ferrimagnetic Co$_2$FeO$_4$, which is expected to contain either amorphous or cubic cobalt-oxide phase contributions for metal ratio retention with regard to the precursor composition of $\mathrm{Fe}:\mathrm{Co} = 1:3$. This result corresponds well to our previous observations for planar Fe-Co deposits \cite{Porrati2015_CoFe}.
\fig{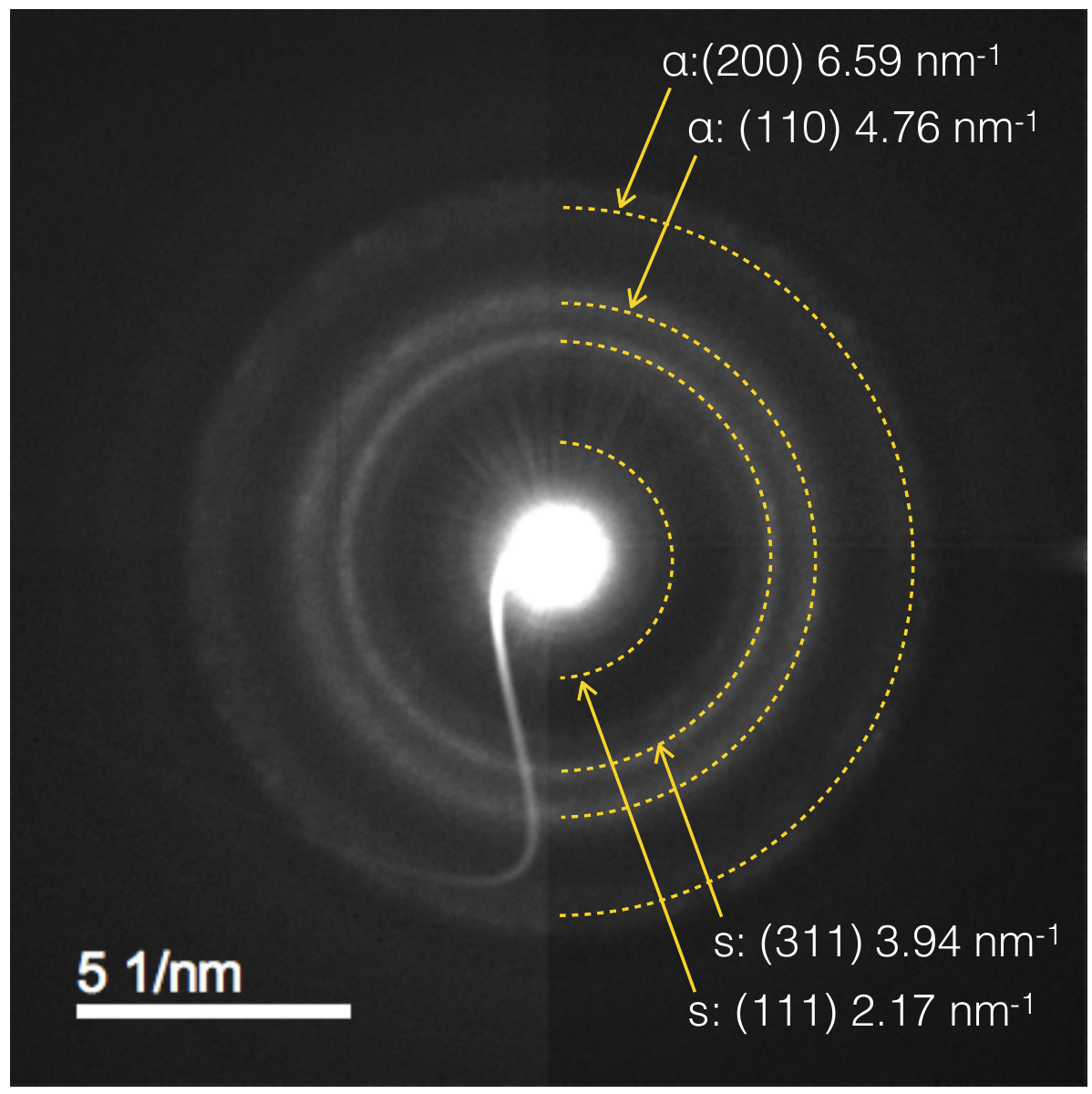}{0.5}{{\bf Electron diffraction.} Selective-area electron diffraction of the region depicted by a green circle in Fig.\,\ref{fig_TEM_EELS}(a). The dotted lines indicate the positions of the diffraction rings associated with the Co-Fe bcc phase ($\alpha$) and the Co$_2$FeO$_4$ spinel phase (s).}{fig_TEM_SAED}

Concluding this part on the microstructural characterization, we are led to consider the following microstructure concerning the magnetic properties of the deposits: a metallic $\alpha$-Co$_3$Fe core is surrounded by a metal-oxide sheath which is a ferrimagnetic spinel phase. Next we turn to the results of the magnetic measurements on the $2\times 2$ arrays of nano-trees and nano-cubes.
%
%
\subsection{Magnetic stray field measurements and macro-spin simulations}
Magnetically frustrated interactions at the three- and four-edge vertices associated with nano-cube and nano-tree structures are expected to lead to non-trivial spatial magnetization profiles and, correspondingly, rather complex distributions of the magnetic stray field vectors. Micro-Hall magnetometry, as sketched in Fig.\,\ref{fig_muHall_and_MS_cubes}(a), is particularly well suited for measuring such stray fields of individual or small arrays of magnetic micro- and nanostructures \cite{Das2010_micro_hall, Pohlit2015_micro_hall_spin_ice_cluster, Pohlit2016_micro_hall_spin_ice_component, Pohlit2016_forc} in a wide range of temperatures and external magnetic fields applied under various angles with respect to the magnetic structures. In Figs.\,\ref{fig_muHall_and_MS_cubes} and \ref{fig_muHall_and_MS_trees} we show stray field measurements of the $2 \times 2$ nano-cube and nano-tree arrays for selected inclination angles of the external magnetic field (see Fig.\,\ref{fig_muHall_and_MS_cubes}(a) for the definition of the angle). The $z$-component of the stray field, $\langle B_z \rangle$, emanating from the magnetic nanostructures is calculated from the measured Hall voltage difference $\Delta V_H = \langle B_z \rangle \cdot I /ne$, where $I$ is the applied current and $ne$ the product of charge carrier density of the GaAs/AlGaAs Hall sensor and the electron charge. $\langle \cdots \rangle$ denotes the average over the active area of the Hall cross and $\Delta V_H$ the {\it in situ} subtraction of the Hall signal of an empty reference cross in a gradiometry setup, see SI and methods section for details.
\fig{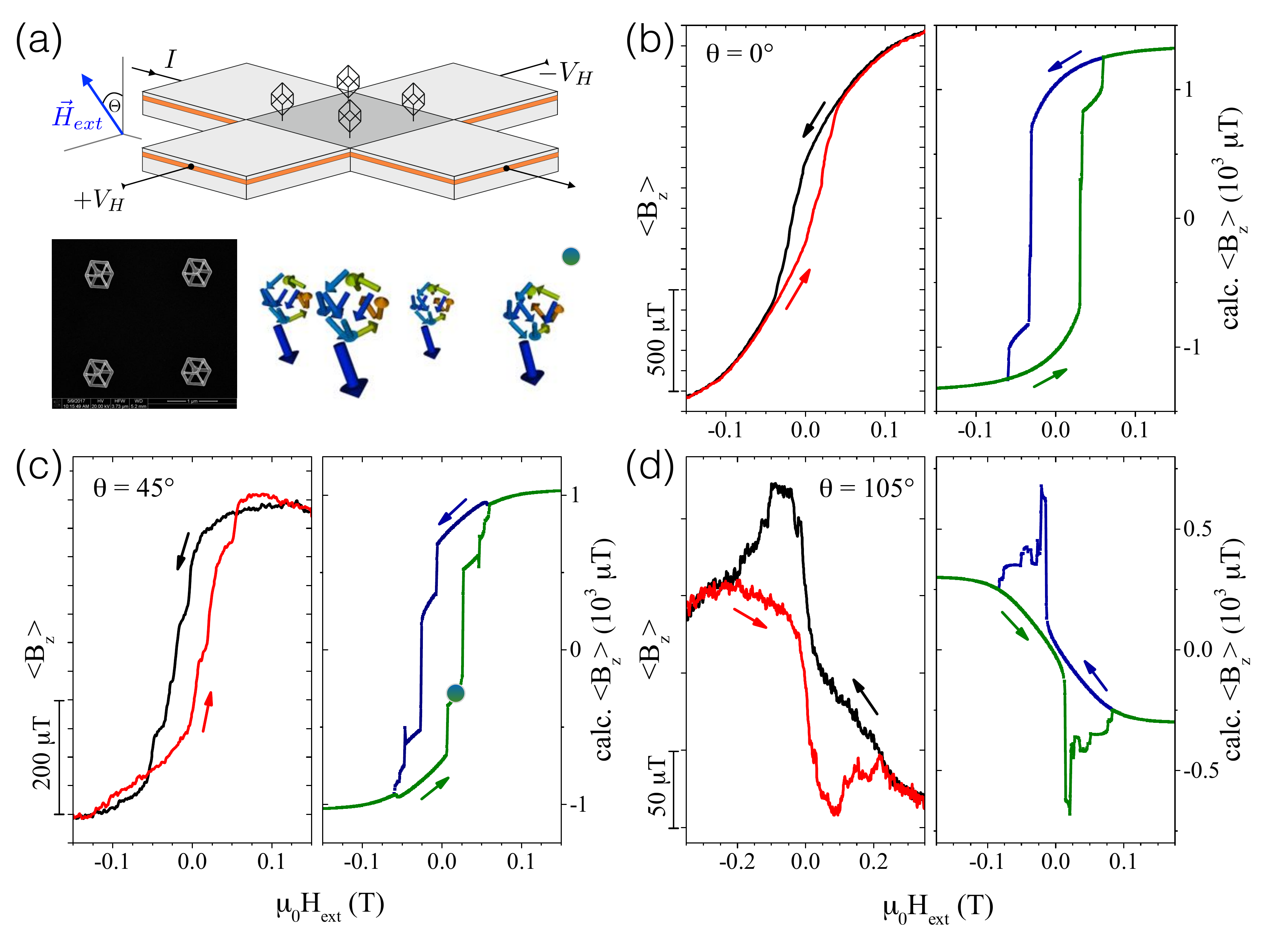}{0.95}{{\bf Micro-Hall stray field measurement results and macro-spin simulations of CoFe nano-cubes.} (a) Sketch of a $2\times 2$ array of CoFe nano-cubes on a $5\times 5\,\mu$m$^2$ Hall cross with the external magnetic field $H_{\rm{ext}}$ applied at an angle $\theta$ relative to the surface normal of the sensor (top). Bottom: SEM micrograph of the nano-cubes (top view) directly written by FEBID on top of the Hall sensor (left) and configuration of the magnetic moments from macro-spin simulations for $\theta = 45^\circ$ (right) shown in (c). The circle marks the position at a positive field in the up-sweep, just before the switching of the stem. (b), (c) and (d) Magnetic stray field hysteresis curves measured at $T = 30$\,K versus calculated stray fields from macro-spin simulations for $T = 0\,$K for a selection of inclination angles $\theta$ equal to $0^\circ$, $45^\circ$ and $105^\circ$, respectively. Arrows indicate the direction of field sweeps.}{fig_muHall_and_MS_cubes}

We first discuss the magnetic nano-cubes. If applied along the $z$-axis ($\theta = 0^\circ$), the external field causes a magnetization reversal which appears to proceed rather continuously, see Fig.\,\ref{fig_muHall_and_MS_cubes}(b). A closer look, however, reveals step-like features in the stray field response. It is therefore instructive to compare the experimental stray field curves with a simple macro-spin approach, where stem and edges of the nano-cubes are represented by a single macro-spin based on the assumption that all microscopic magnetic moments point in the same direction and rotate collectively. Since such an approach omits the existence of form anisotropy this has to be modeled by an additional uniaxial anisotropy for each macro-spin, see methods section for more details and the parameters used, see the SI for a discussion of the deviations between the measured and simulated absolute stray field signals. The model is lattice-based and mesh-free, which makes it very efficient for computing the mutual dipolar interactions of stems and edges within the nano-cube and -tree arrays. The almost vertical decrease or increase of $\langle B_z \rangle$ of the simulated up- and down-sweep curves, respectively, at $\theta = 0^\circ$ shown in the right panel of Fig.\,\ref{fig_muHall_and_MS_cubes}(b) are associated with the flipping of the four stems whereas the smaller steps are caused by the almost simultaneous flipping and canting of the edge macro-spins. Although the qualitative agreement is satisfying, the more rounded loop with smaller area and coercive field observed in the experiment points to a non-uniform magnetization switching of the stems dominated by multi-domain switching events.

A pronounced step-like switching behavior and better agreement of the measurements with the model is observed for a tilt angle of $\theta = +45^\circ$ as shown in Fig.\,\ref{fig_muHall_and_MS_cubes}(c). Again, the large steps are associated with the flipping of the stems. Smaller steps and the finite slopes in between are connected with the flipping of edge spins and the rotation of their magnetization direction towards the external field. The closer the direction of the external field is to the anisotropy axis of an edge spin the larger is the coercive field resulting in the observed stair-case shape of the hysteresis.

A remarkable qualitative correspondence of the measured and simulated curves is seen for the complex and strongly pinched hysteresis loop at $\theta = +105^\circ$ shown in Fig.\,\ref{fig_muHall_and_MS_cubes}(d), where $H_{ext}$ is almost perpendicular to the anisotropy axis of the stems. Upon lowering the absolute value of the field, e.g.\ from negative saturation, all macro-spins relax towards their anisotropy axes. However, when approaching $H_{ext} = 0$, the edge spins are not in the lowest energy state but their total moment has a large component parallel to the external field axis whereas the stem spins point upwards parallel to the anisotropy axis. This remains the case even for small positive fields and causes the stem spins to suddenly rotate by $180^\circ$ then pointing downwards. This causes the sharp peaks in the hysteresis which are observed in the measurements as well, however smeared out due to finite temperature and because the magnetization reversal mechanism of stem and edges are more complicated than the coherent rotation assumed in a single-domain macro-spin model. Nevertheless, although idealized, macro-spin simulations allow for identifying the relevant switching scenarios that occur for different inclination angles.
\fig{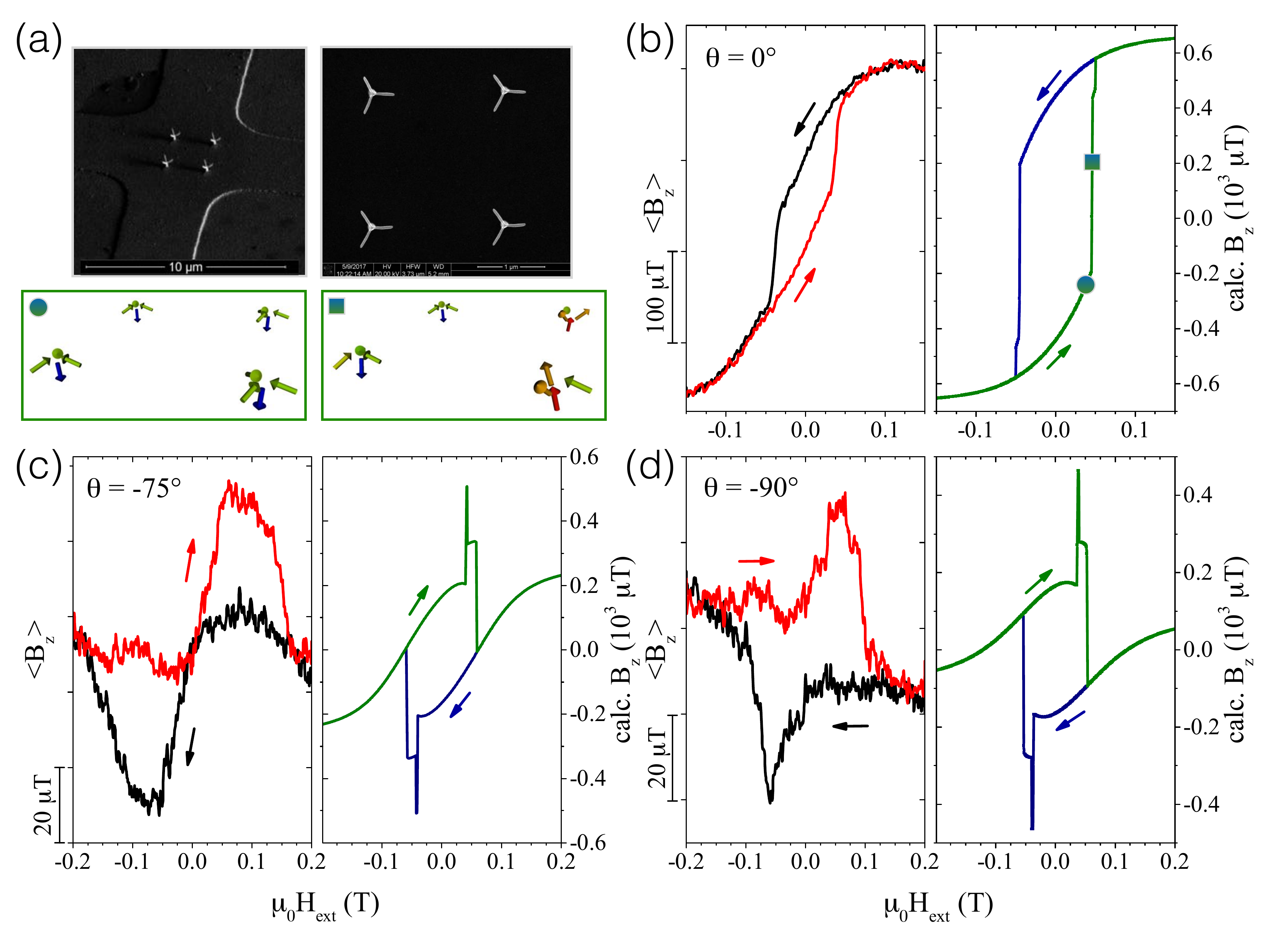}{0.95}{{\bf Micro-Hall stray field measurement results and macro-spin simulations for nano-trees.} (a) (a) Top: SEM micrograph of a $2\times 2$ array of CoFe nano-trees on a $5\times 5\,\mu$m$^2$ Hall cross deposited by FEBID (left) and top view of the nano-trees (right). Bottom: Simulations showing the macro-spin configuration at $\theta = 0^\circ$ for the up-sweep at two positive fields just before and at the switching of the stems' magnetization marked by circle and square in (b). (b), (c) and (d) Comparison of the magnetic stray fields $\langle B_z \rangle$ at $T = 30\,K$ and macro-spin calculations for $T = 0\,$K and selected angles $0^\circ$, $-75^\circ$ and $-90^\circ$, respectively. Arrows indicate the direction of field sweeps.}{fig_muHall_and_MS_trees}

Next, we focus on the comparison of the measured nano-trees' magnetization reversal with macro-spin simulations followed by a more sophisticated theoretical investigation based on micromagnetic simulations (see next section). As for the nano-cubes discussed above, the angular dependence of the measured stray fields, exemplarily shown for three angles $\theta = 0^\circ$, $\theta = -75^\circ$ and $\theta = -90^\circ$ shown in Fig.\,\ref{fig_muHall_and_MS_trees}(b), (c) and (d), respectively, demonstrates a rich variety of reversal processes determined by the field angle with respect to the stem and edges of the 3D nano-trees. If the external field $H_{ext}$ is applied parallel to the stem ($\theta = 0^\circ$), the continuous progression of the stray field upon decreasing $H_{ext}$ from saturation indicates a gradual rotational change of the magnetization direction followed by a switching process comprising a substantial part of the sample volume. The experimental hysteresis is qualitatively reproduced by the macro-spin model, which identifies the gradual change in magnetization with an umbrella-like magnetic canting and rotation of the edges towards the direction of the applied field, while a fast switching sequence of the stems accounts for the sudden change of the magnetization, see the sequence of macro-spin configurations in the lower panel of Fig.\,\ref{fig_muHall_and_MS_trees}(a) corresponding to two field values marked in the hysteresis loop of Fig.\,\ref{fig_muHall_and_MS_trees}(b).

At an incident field angle of $\theta = -75^\circ$ the field is nearly parallel to one of the edges. Here, a distinctly different magnetic hysteresis which narrows at zero applied field and exhibits a broad minimum and maximum at negative and positive fields for the down- and up-sweep curves, respectively, is observed. This behavior is only partly reproduced by the macro-spin model. The hysteresis loop shows maxima in the field cycle, but no narrowing of the loop occurs around zero field. Such a behavior requires a more finely grained effective spin model and is well reproduced by micromagnetic simulations considering a non-uniform reversal mechanism in the presence of the thin metal-oxide sheath, as shown in Fig.\,\ref{fig_muHall_and_MM_trees} below. In contrast, certain features of the stray field hysteresis for $\theta = -90^\circ$ being less pinched and exhibiting sharper peaks are found in the macro-spin model. The pronounced peaks correspond to the rotation of the stems' magnetization relative to the edges. At large fields, the magnetization of the edges and the stem are aligned along the field direction. Compensating stray fields from oblique magnetization angles of edges and stems lead to the small kinks near the peaks at smaller fields. A stronger compensation, which would reproduce the experimentally observed narrowing over a wide range of external fields at $\theta = -75^\circ$ and $\theta = -90^\circ$ is not found within the macro-spin model. Indeed, micromagnetic calculations, which are described in the next section, show a much more complex switching behavior beyond the limits of a  fixed magnetic moment and a single axis anisotropy assumed in a macro-spin model. 
%
%
\subsection{Micromagnetic simulations}
Quite generally, micromagnetic simulations are invaluable for obtaining a deeper understanding of hysteresis effects by visualization of the magnetization reversal process on a microscopic scale \cite{Fidler2000_micromagnetism}. At the same time, they are computationally much more demanding than macro-spin simulations, which still limits their application depending on the size of the simulation volume. However, using the parallel computing power of high-end graphics cards has led to about an order of magnitude faster code execution (see, e.\,g.\ \cite{Vansteenkiste2014_mumax3}) and the development of multi-scale and multi-physics of micromagnetic solvers is already foreseeable \cite{Kruglyag2010_magnonics_review, Dvornik2013_micromagnetic_simulations}. Here we present results of micromagnetic simulations for the reversal process of the nano-trees in order to provide insight into the reasons for the discrepancy between the micro-Hall data and the macro-spin simulations. We also shed light onto the importance of taking the different magnetic behavior of the near-surface oxide into account. We focus on the nano-trees, because for these, the discrepancies compared to macro-spin simulations are most pronounced and they can still be treated by micromagnetic simulations if a core-shell structure consisting of metallic CoFe$_3$ (core) and a ferrimagnetic spinel phase of Co$_2$FeO$_4$ (shell) is assumed. Details on the simulation parameters are given in the methods section. For results of micromagnetic simulations of the nano-cubes we refer to the SI.

In Fig.\,\ref{fig_muHall_and_MM_trees}(b), (c) and (d) we present the results of the micromagnetic simulations at the same angles which have been shown before. We discuss two different material composition scenarios. The first scenario assumes an all-metal Co$_3$Fe nano-tree, wheres the second scenario assumes an oxide shell of the ferrimagnetic spinel phase Co$_2$FeO$_4$ covering an all-metal Co$_3$Fe core, as is schematically indicated in Fig.\,\ref{fig_muHall_and_MM_trees}(a). For the spinel phase, the saturation magnetization is assumed to be a factor of $10$ smaller than that of the Co$_3$Fe core (see methods section for details where it is also explained how the average stray field was calculated.). 
\fig{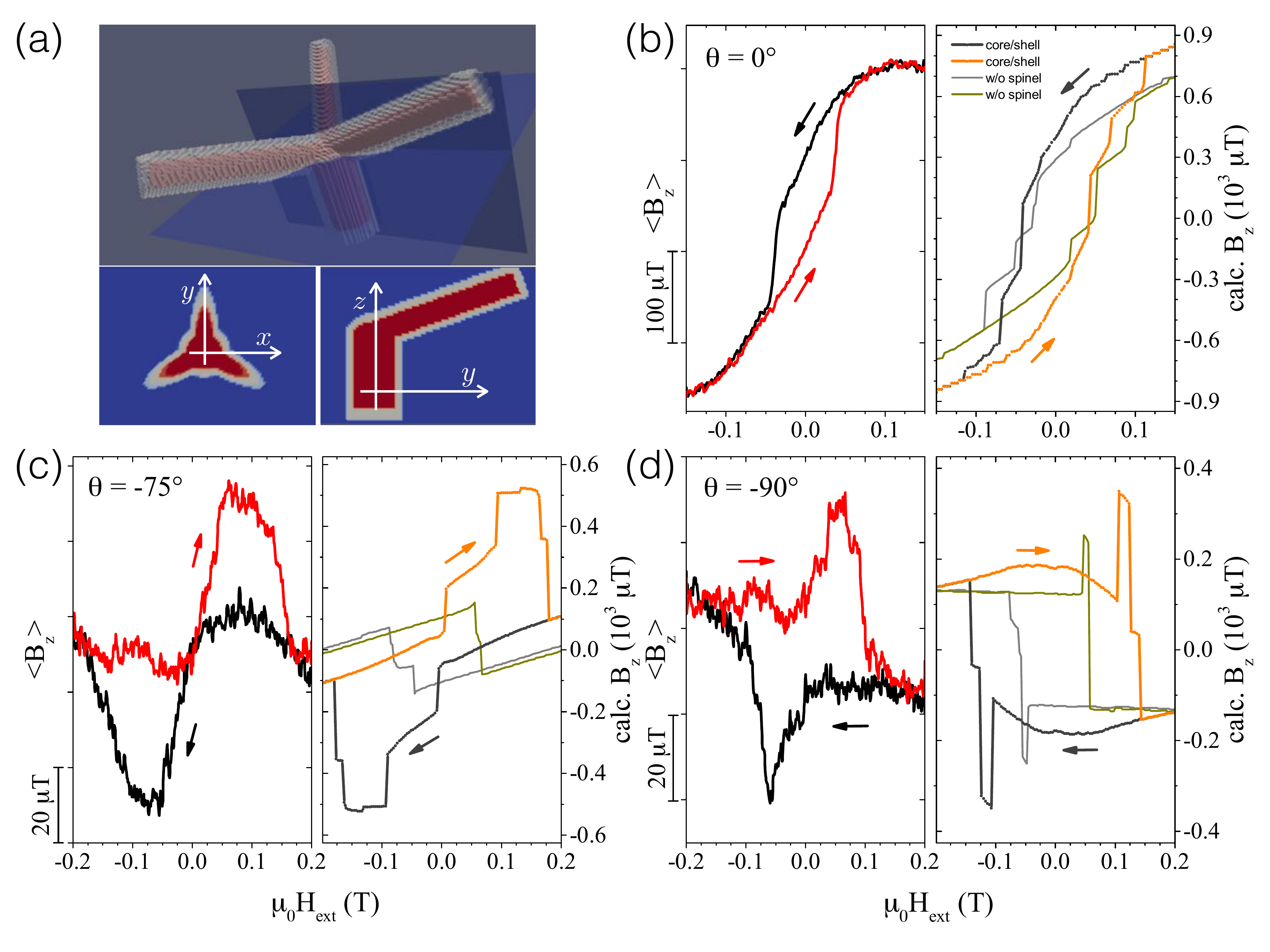}{0.95}{{\bf Micro-Hall stray field measurement results and micromagnetic simulations for nano-trees.} (a) 3D-view and cross sections of nano-tree assuming a Co$_3$Fe core (red) and Co$_2$FeO$_4$ spinel shell (gray), according to scenario 2 (core-shell structure). (b), (c) and (d) Comparison of the magnetic stray fields $\langle B_z \rangle$ at $T = 30\,K$ and micromagnetic simulations for $T = 0\,$K with or without core-shell structure (as indicated) for selected angles $0^\circ$, $-75^\circ$ and $-90^\circ$, respectively. Arrows indicate the direction of field sweeps.}{fig_muHall_and_MM_trees}
For the all-metal micromagnetic model we find the same qualitative behavior as in the macro-spin simulations. The overall correspondence with the micro-Hall data for $\theta = 0^\circ$ is good and a reasonable agreement can be stated for $-90^\circ$. This indicates that the macro-spin model catches the main features of the magnetization reversal processes in these cases, however see also the SI for a more detailed presentation of the spatial magnetization distribution as obtained from the micromagnetic simulations. In contrast to this, for $\theta = -75^\circ$ the micro-Hall data are not reproduced by the all-metal micromagnetic or macro-spin simulations. However, if the core-shell structure with a spinel shell of reduced saturation magnetization is taken into account, we find for all angles a very good correspondence with the data. In particular, the pinched hysteresis form for $\theta = -75^\circ$ is very well reproduced, and the shapes of the stray-field hysteresis for $\theta = 0^\circ$ and $\theta = -90^\circ$ are also very similar to the ones obtained by micro-Hall magnetometry. In addition, we observe that the coercive fields correspond quite well to the measured values. With regard to the remaining differences one has to take into account that our micromagnetic simulations assume $0\,$K, whereas the micro-Hall data shown here were taken at $30\,$K.
%
%
\section{Discussion}
We have shown that the main features of the magnetization switching of the 3D nano-architectures, as monitored by micro-Hall magnetometry, are already quite well reproduced by a fast and scalable macro-spin approach. If complemented by carefully designed micromagnetic simulations, the correspondence becomes very satisfying also in those cases for which the macro-spin model is less successful. In addition we note that from our micromagnetic simulations it becomes also quite apparent that for the presented case of frustrated interactions through a vertex, which is magnetic itself, the magnetization distribution inside the magnetic 3D structures can be rather complex (see SI for details). In view of a prospective application of the presented building blocks towards 3D artificial spin-ice systems it may be desirable to reduce this level of complexity in the magnetization distributions. For such arrays, micromagnetic simulations will not be feasible, and it will be exceedingly difficult to acquire a satisfying understanding of all details of the array's switching behavior. Thus, we consider replacing the vertices in the nano-elements by non-magnetic material and demonstrating that the nano-elements can be arranged in 3D array structures by our FEBID approach. Fig.\,\ref{fig_array_FeCoPt_SEM} illustrates first results in these directions for the nano-tree geometry. Figure parts (a) and (b) show top and tilted views of a 3D nano-tree array employing again the precursor HCo$_3$Fe(CO)$_{12}$. In this case the FEBID writing strategy has to be carefully adapted to compensate for precursor gas flux shadowing effects \cite{Winkler2014_gas_flux_influence} and anisotropic growth or proximity bending of growing nano-elements\cite{Winkler2017_3D_plasmonic}. If done properly, the nano-trees' shape reproducibility and the placement accuracy are very good, as shown here. In figure parts (c) and (d) we demonstrate that it is furthermore possible to replace the vertex segment in the nano-tree by non-magnetic material, in our case nano-granular Pt, using Me$_3$CpMePt(IV) as precursor.
\fig{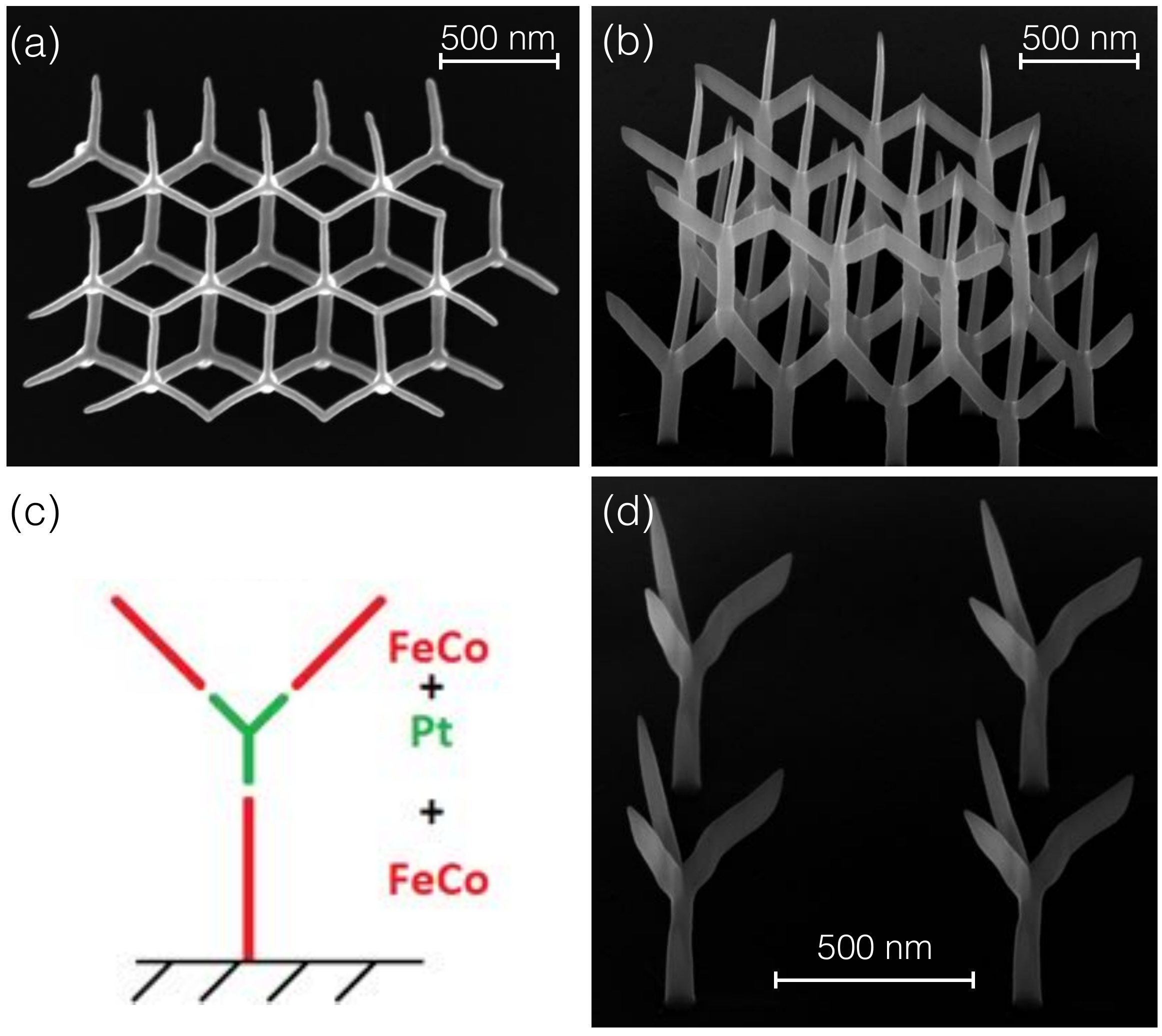}{0.95}{{\bf 3D array of nano-trees and nano-tree with non-magnetic vertex.} (a) SEM top view of 3D nano-tree array fabricated by FEBID with the precursor HCo$_3$Fe(CO)$_{12}$. (b) Tilted view of the same array. (c) Schematic of nano-tree for which the vertex part is replaced by a non-magnetic segment consisting of nano-granular platinum Pt(C). (d) SEM image of a $2\times 2$-array of nano-trees with FeCo stems and edges and non-magnetic Pt(C) vertex segment.}{fig_array_FeCoPt_SEM}
%
%
%
\section{Conclusion}
We have demonstrated that FEBID provides a powerful and flexible way to realize free-form magnetic 3D nano-elements and arrays of such structures. Our TEM-based characterization clearly indicates that HCo$_3$Fe(CO)$_{12}$ is a particularly well-suited precursor, as it leads to 3D deposits with nominally pure metallic character under beam conditions which are perfectly suitable for high-resolution structuring. The microstructure and composition analysis allowed us to pinpoint the essential features for suitable micromagnetic modeling (core-shell structure) and we were able to reproduce the most important stray field effects, as observed by micro-Hall magnetometry. In macro-spin model calculations, which are favorable because of their scaling behavior towards larger arrays, several observations in the switching behavior of the 3D nano-magnets can already be well reproduced. Replacing the vertex segment of the nano-elements by a non-magnetic material and the arrangement of nano-elements in 3D arrays was successfully demonstrated employing the FEBID approach. This will be of advantage for future work on 3D artificial spin-ice which is but one example of various other possible application fields of 3D magnetic FEBID structures on the single-element and array basis. Mesoscopic 3D arrays using this approach may allow for experimentally realizing and studying classical Ising or Heisenberg model systems \cite{Melko2001_dipolar_spin_ice}. A chiral geometry of the nano-elements in 3D arrays is of high interest with regard to magneto-optical properties (see, e.\,g.\ \cite{Eslami2014_chiral_nanomagnets}). Finally, single 3D nano-magnetic elements with increasing geometrical complexity are conceivable and will pave a new way for bringing topology-design elements into micromagnetic research.
%
%
\section{Methods}
%
%
\subsection{FEBID}
Samples were fabricated using a dual beam SEM/FIB (FEI, Nova NanoLab\,600), equipped with a Schottky electron emitter operating at a base pressure of about $2\times 10^{-7}\,$mbar. The  precursors HCo$_3$Fe(CO)$_{12}$ and Me$_3$CpMePt(IV) (Me: methyl, Cp: cyclopentadienyl) were injected in the SEM by means of a capillary with an inner diameter of 0.5\,mm. The distance capillary-surface was about $100\,\mu$m and the tilting angle of the injectors was $50^\circ$. The crucible temperature of the gas injection system (GIS) was set to $65^\circ$C and $45^\circ$C for the Fe-Co and Pt precursor, respectively. The electron beam parameters used during deposition were 20\,keV for the acceleration voltage and 13\,pA for the beam current. The dwell time was set after optimization of the 3D growth to $1\,$ms. The pitches depend on the inclination angle of the 3D structures and have to be adapted to the precursor and gas flow conditions \cite{Fowlkes2016_3D_FEBID}. Concerning the synthesis of the Co-Fe precursor we refer to \cite{Porrati2015_CoFe}.
%
%
\subsection{TEM}
TEM investigations were carried out on a Tecnai F20 from FEI with a Schottky Field Emitter operating at $200\,$kV. Images were taken with a post-column energy filter (Gatan Imaging Filter, GIF) using an energy slit of $10\,$eV. The images were recorded zero-loss filtered (i.\,e., elastically scattered electrons only) on a 2K charge coupled device. For the image recording and processing (Fourier transformation) the software DigitalMicrograph from Gatan was used.
%
%
\subsection{Micro-Hall magnetometry}
The basic principle of operation of micro-Hall magnetometry is schematically sketched in Fig.\,\ref{fig_muHall_and_MS_cubes}(a). When a magnetic sample is placed on top of the sensor, in first approximation, the measured Hall voltage, $V_H$, is proportional to the $z$-component of the sample's magnetic stray field averaged over the active area of the Hall cross, $\langle B_z \rangle$, via $V_H = 1/ne \cdot I \cdot \langle B_z \rangle$. Here, $n$ denotes the charge carrier concentration, $I$ is the applied current, and $e$ is the electron charge. A more detailed account, including a discussion of background subtraction and magnitude of the measured stray fields, is given in the SI.

The homebuilt Hall sensor is fabricated from an MBE-grown AlGaAs/GaAs heterostructure hosting the two-dimensional electron gas (2DEG) as the sensitive layer. In a first step, standard UV-lithography followed by wet chemical etching is employed to form six adjacent Hall-crosses of $5 \times 5\,\mu{\rm m}^2$ nominal size. Then the sensor structure is electrically contacted by annealed AuGe/Ni contact pads and gold wire bonding. Subsequently the sensor is covered with a Cr/Au top-gate which also serves as substrate for the 3D nano-cubes and -trees directly written by FEBID, see the SEM micrograph shown in Fig.\,$1$(b) in the SI. The gate is grounded during the measurements. The sensor can be operated in a wide magnetic field and temperature range, but is optimized for $4.2\,K \lesssim T \lesssim 100$\,K.

 After writing of the 3D magnetic nanostructures the Hall sensor has been transferred in less than one hour to a cryogenic system equipped with a superconducting solenoid essentially free of magnetic flux jumps. The sample can be rotated with respect to the applied magnetic field, where $\theta = 0^\circ$ for field angles perpendicular to the sensor plane.
%
%
%
\subsection{Macro-spin simulations}
Each element of the nano-trees and nano-cubes is modeled by a single macro-spin. The form anisotropy of each macro-spin is accounted for by a single uniaxial anisotropy. For the determination of the strength of this uniaxial anisotropy we assume that each edge and stem can be associated with a prolate ellipsoid. This allows us to calculate the anisotropy constant by means of the Stoner-Wohlfarth model \cite{Pohlit2016_micro_hall}. Assuming an average magnetization of $1500\,$kA/m the corresponding uniaxial anisotropy values lead to coercive fields much higher than the experimental ones. Therefore, we have kept the calculated ratio between the stem and edge anisotropy of 1:1.3 and fitted the anisotropy constant to the experimental coercive  field at an inclination angle of $0^\circ$. With a stem anisotropy constant of $285,520\,$eV and a edge anisotropy constant of $336,076\,$eV significant features of the experimental stray  field hysteresis curves can be reproduced. This allows us to identify the underlying switching behavior of the different elements.

In order to study the dynamics of our macro-spin model we numerically solve the stochastic Landau-Lifshitz--Gilbert equation
\begin{equation}
\frac{\partial\vec{S}_i}{\partial t} =-\gamma \left( \frac{\partial H}{\partial \vec{S}_i}- \vec{f}_i \right) \times \vec{S}_i - \alpha \gamma \left( \frac{\partial H}{\partial \vec{S}_i} \times \vec{S}_i \right) \times \vec{S}_i \,. 
\label{eq_LandauLifshitz} 
\end{equation}
Eq.\,\ref{eq_LandauLifshitz} describes the motion for a macro-spin $\vec{S}_i$ of unit length at site $i$ caused by an effective field $\vec{H}_{\rm{eff}}$ which is generated by the interactions given in the Hamiltonian Eq.\,\ref{eq_ExtendedHeisenberg} below. In Eq.\,\ref{eq_LandauLifshitz} $\gamma$ denotes the gyromagnetic ratio and $\alpha$ a phenomenological damping factor in front of the so-called Landau-Lifshitz damping term that drives the macro-spin towards a full alignment with the effective field $\vec{H}_{\rm{eff}}$, i.\,e.\ a local or global minimum configuration. Eq.\,\ref{eq_LandauLifshitz} also allows to study finite-temperature effects by setting
\begin{equation}
\left< f_i^\alpha(t) f_j^\beta(t')\right> = \epsilon^2 \delta_{ij}
\delta_{\alpha \beta} \delta(t-t'),
\label{eq_fluctuations}
\end{equation}
where $\epsilon$ is the amplitude of the fluctuations and defined by
\begin{equation}
\epsilon^2 = 2\lambda k_B T \,.
\label{eq_amp_fluc}
\end{equation}
The evolution of the trajectories $\vec{S}_i(t)$ obtained by solving Eq.\,\ref{eq_LandauLifshitz} and \ref{eq_fluctuations} leads to the stationary Gibbs distribution for a given temperature $T$. The macro-spin calculations are used to study the dynamic switching behavior of the nano elements. As the temperature of such non-equilibrium processes within a macro-spin model does not correspond to a physical temperature, we have performed the macro-spin simulations at $T = 0\,$K.

The Hamiltonian describing the interactions among all macro-spins is given by
\begin{equation}
H = - D \sum_{i=1} \left( \mu_{\rm{i}} \vec{S}_{i} \cdot \vec{e}_i \right)^2
 -\frac{\mu_0 }{4 \pi} \sum_{i<j} \mu_{\rm{i}} \mu_{\rm{j}} \frac{3  \left( \vec{S}_{i} \cdot \vec{e}_{i,j} \right) \left( \vec{e}_{i,j} \cdot  \vec{S}_{j} \right) - \vec{S}_i \cdot \vec{S}_j}{r^{3}_{i,j}} 
 -\vec{B}_{\rm{ext}} \cdot  \sum_{i=1} \mu_{\rm{i}}  \vec{S}_{i} 
\label{eq_ExtendedHeisenberg}
\end{equation}
Here, the first term describes a uniaxial anisotropy, where $D$ is the anisotropy constant and $\vec{e}_i$ is the unit vector pointing into the anisotropy direction. For $D < 0$ this term describes an easy-axis anisotropy. The second term is the dipole-dipole interaction, where $\mu_i$ describes the effective magnetic moment per macro-spin for the stem or the edge, respectively. The direction between two interacting macro-spins is given by the unit vector $\vec{e}_{i,j}$ and the distance is given by $r_{ij}$. The last term in Eq.\,\ref{eq_ExtendedHeisenberg} is the Zeeman term which describes the interaction of the macro-spins with the external magnetic field $\vec{B}_{\rm{ext}}$.

We obtained our hysteresis curves by linearly ramping up and down the external field in the range of $-0.2\,\rm{T} \leq B_{\rm{ext}} \leq 0.2\,\rm{T}$ using $10^6$ time steps of length $\Delta t = 10\,$ps. Using a damping constant of $\alpha = 0.3$ such calculations are very fast and just need minutes on a single core of a common computer processor. The calculation of the stray fields $\langle B_z\rangle$ was done in the following way. (1) The positions of the four nano-trees and nano-cubes on the Hall sensor area where determined from SEM images. (2) The cumulated stray field contributions of all macro-spins of the nano-tree/cube have been averaged over $420\times 420$ positions in the $xy$-plane of the sensor array area (roughly $5\times 5\,\mu$m$^2$) at the $z$-position of the 2DEG $115\,$nm below the substrate surface.
%
%
\subsection{Micromagnetic simulations}
Zero temperature micromagnetic simulations were performed by numerically solving Eq.\,\ref{eq_LandauLifshitz} for a single nano-tree consisting of Co$_3$Fe (scenario 1), Co$_3$Fe / Co$_2$FeO$_4$ core/shell-structure (scenario 2) or a single nano-cube of Co$_3$Fe (see SI for nano-cube). We used the GPU-accelerated micromagnetic simulation program MuMax3 \cite{Vansteenkiste2014_mumax3} running on a Linux notebook with Intel Core i7-7700HQ processor, 32\,GB random access memory and NVidia GeForce GTX 1060 graphics card. Using cubic voxels of edge length $5\,$nm for the finite difference discretization in MuMax3 the simulations for a typical external field cycle $-0.2\,\rm{T} \leq \mu_0H_{\rm{ext}} \leq 0.2\,\rm{T}$ at a step size of 0.0033\,T took about 60\,hrs for one nano-tree. Simulations for the nano-cube with core/shell structure were not attempted, as the simulation volume and voxel number was expected to lead to simulation time of more than 260\,hrs per external field cycle. Nano-cube simulations assuming full metallic Co$_3$Fe as material were performed within 6\,hrs for a typical field cycle. The simulation parameters were chosen as follows:
\begin{description}
\item[Nano-tree] Geometrical dimensions: stem diameter $D_s = 119\,$nm (cylindrical) and length $L_s = 185\,$nm, edge diameters $D_{b,1} = 80\,$nm and $D_{b,2} = 64\,$nm (elliptical) at a length of $L_b = 340\,$nm, thickness of oxide spinel shell $t = 2\lambda$ (scenario 2). Material parameters: the saturation magnetization of the shell was set to $M_S^{(\rm{Co_2FeO_4})} = 1.15\times 10^5\,$A/m using experimental data from \cite{Muthuselvam2009_spinel}. For the Co$_3$Fe-core we used $M_S^{(\rm{Co_3Fe})} = 1.5\times 10^6\,$A/m and the exchange constant $A = 1.4\times 10^{-11}\,$J/m by averaging the respective value for Fe and Co \cite{Porrati2004_diagram_states_Fe, Pohlit2016_micro_hall}. As we could not find a reference for the exchange constant of the spinel we used the same exchange constant as for the core material.
\item[Nano-cube] Geometrical dimensions: stem diameter $D_s = 119\,$nm (cylindrical) and length $L_s = 185\,$nm, edge diameters $D_b = 62\,$nm (assumed cylindrical) at a length of $L_b = 340\,$nm. Material parameters: the saturation magnetization was set to $M_S^{(\rm{Co_3Fe})} = 1.5\times 10^6\,$A/m and the exchange constant to $A = 1.4\times 10^{-11}\,$J/m.
\end{description}
The nano-grain microstructure of the deposits leads to an averaging of the magnetic anisotropy, which is why we have omitted any anisotropy energy contributions in our simulations. In order to guarantee sufficiently fast convergence we set the damping parameter $\alpha$ to $0.3$, used the full relaxation of MuMax3 \cite{Vansteenkiste2014_mumax3} at the initial field value and then the conjugated gradient method for quicker convergence at all other field values of each cycle with a stop criterion of $10^{-6}$.

The simulation data on the orientation of the magnetic moment vectors within the volume elements of the nano-tree or nano-cube for each external field was used to calculate the stray field $\langle B_z\rangle$ at the sensor layer in the following way. (1) The positions of the four nano-trees and nano-cubes on the Hall sensor area where determined from SEM images. (2) For each volume element of the nano-tree/cube the associated simulated magnetic moment was used to calculate the corresponding dipolar stray field. The stray field contributions of all moments of the nano-tree/cube set to one of the four positions were averaged over $28\times 28$ positions of the sensor array area (roughly $5\times 5\,\mu$m$^2$) in the $xy$-plane at the $z$-position of the 2DEG $115\,$nm below the substrate surface. (3) The resulting four averaged stray fields were added to obtain the full averaged stray field of the four nano-trees/cubes.
%
%
\bibliographystyle{unsrt}
\bibliography{magnetic3d_submitted_rev1}
%
%
\section{Acknowledgements}
M.~H.\ and L.~K.\ acknowledge financial support by the Deutsche Forschungsgemeinschaft (DFG) through the Collaborative Research Centre SFB/TR\,49. I.~S.\ acknowledges financial support from the Swedish Government Strategic Research Area in Materials Science on Functional Materials at Link\"oping University (Faculty Grant SFOMatLiU No 2009 00971). M.~A.~M.\ acknowledges financial support from the Faculty of Materials Engineering, University of Babylon, Babylon, Iraq, and from the Deutscher Akademischer Austauschdienst (DAAD) within the doctoral program for research studies in Germany. The MBE-grown high-mobility wafer material that was used to build the Hall magnetometer was kindly provided by Dr. J\"urgen Weis, Max-Planck-Institute for Solid State Research, Stuttgart, Germany. H.~P.\ thanks Prof.\ Ferdinand Hofer, Prof.\ Werner Grogger and Prof.\ Gerald Kotleitner for support. This work was conducted within the framework of the COST Action CM1301 (CELINA).
%
%
\section{Author contributions}
M.~H.\ and J.~M.\ devised the project. L.~K.\ performed the FEBID process with support by R.~W.\ and H.~P.. M.~A.~M.\ and J.~P.\ performed the micro-Hall measurements. M.~P.\ prepared the micro-Hall sensor and supported the stray-field data analysis. I.~S.\ and C.~S.\ performed the macro-spin simulations. Micromagnetic simulations were done by M.~H.. S.~B.\ synthesized the precursor. C.~G.\ did the TEM experiments and analysis, supported by H.~P.. M.~H.\ and J.~M.\ wrote the manuscript with input from all authors.
%
%
\section{Additional information}
None
%
%
\section{Competing financial interests}
The authors declare no competing financial interests.
%
%
\newpage
\section*{Supplementary Information}
%
%
\section{Micro-Hall magnetometry}
%
%
\subsection{Principle of measurement}
\begin{figure}
\centering
\includegraphics[width=0.7\textwidth]{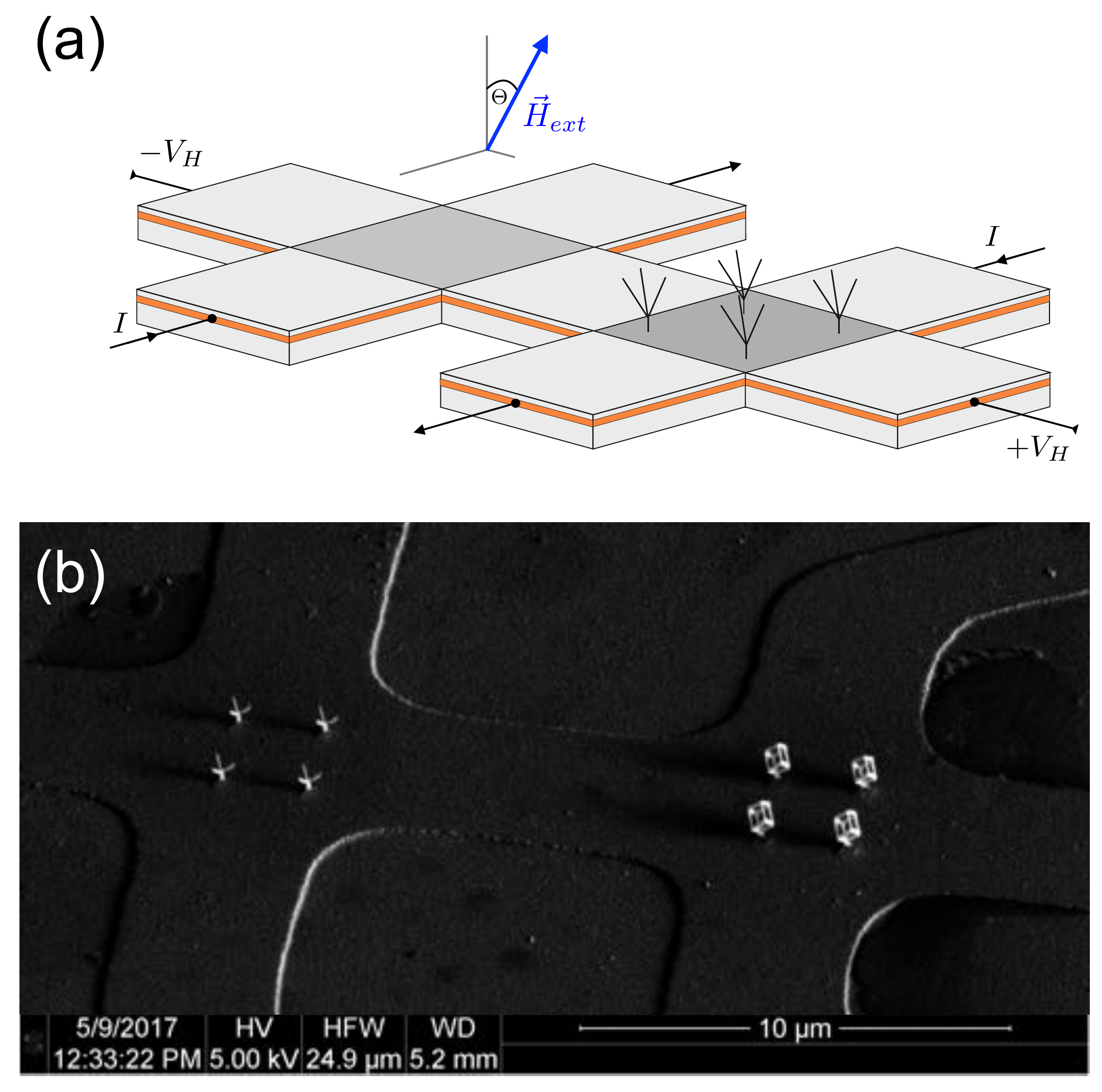} 
\caption{{\bf Micro-Hall gradiometry setup} (a) Schematic of the Hall gradiometry technique, which allows for an {\it in situ} background correction by subtracting the signal which is linear in $H_{ext}$ of an empty reference cross. The definition of the angle of the applied (b) SEM micrograph of the two adjacent Hall crosses accommodating $2 \times 2$ arrays of CoFe nano-trees and -cubes. The empty reference cross to the left is not shown.}
\label{fig_SI_gradiometry_scheme}
\end{figure}

The magnetic stray field of a sample -- which is directly linked to its magnetization \cite{Pohlit2016_micro_hall} -- is measured by detecting the Hall voltage $V_{\rm{H}}$ generated in the sensor plane formed by the 2DEG at the interface of the AlGaAs/GaAs heterostructure, see the schematics in SI-Fig.\,\ref{fig_SI_gradiometry_scheme}(a). In first approximation, the detected $z$-component of the stray field averaged over the active area of the Hall-cross $\langle B_{\rm{z}} \rangle$ is directly proportional to the measured Hall voltage $V_H$. Since the integrated stray field of the arrays of nano-cubes or -trees grown on top of the Hall sensor, see SI-Fig.\,\ref{fig_SI_gradiometry_scheme}(b), is more than two orders of magnitude smaller than the applied external field $\mu_0 H_{ext}$, a so-called gradiometry measurement is performed, where the large signal, which is linear in $\mu_0 H_{ext}$ is cancelled {\it in situ} by applying opposite currents across two Hall crosses, one decorated with magnetic particles and one empty, see Fig.\,\ref{fig_SI_gradiometry_scheme}(a). Then, the stray field contribution of the magnetic nanostructures is given by  
\begin{equation}
\Delta V_H  = \frac{1}{n e} \cdot I \cdot \langle B_z \rangle \,,
\label{eq_1}
\end{equation} 
where $n = 3.4 \times10^{11}\,\rm{cm^{-2}}$ denotes the carrier density of the sensor and $I$ the applied currents, which have been $I = 2.5\,{\rm \mu A}$. Data of $\langle B_z \rangle$ vs.\ $\mu_0H_{ext}$ shown in this work have been corrected by subtracting a small linear background caused by slight differences between the two crosses in the gradiometry setup.
%
%
\subsection{Stray field magnitude in measurement and simulation}
Figures 4, 5 and 6 of the main paper compare the magnetic stray fields measured by micro-Hall magnetometry with macro-spin and micromagnetic simulations. In all cases, the magnitudes of the simulated $z$-components of the stray fields are larger than the experimental values for $\langle B_z \rangle$. For example, the measured remanent field for the nano-cubes at $\theta = 45^\circ$ shown in Fig.\,4 is a factor of 4.5 smaller than the values simulated in the macro-spin model. Here, we briefly discuss possible reasons.

In \cite{Peeters2002} it is pointed out that introducing an effective Hall coefficient and a Hall response function $F_H(x,y)$ in the ballistic and diffusive transport regime, respectively, allows for taking account of the specific geometry of the Hall bar, such as circular corners and asymmetry in the probes. At $T = 30$\,K, where the measurements shown in this work have been performed, the transport properties of the 2DEG are considered in the diffusive regime, and therefore: 
\begin{equation}\
\langle B_z(x,y) \rangle = \frac{\Delta V_H \cdot n e}{I} =  \frac{\sum\limits_{i = 1}^4\int_A {\rm d}x {\rm d}y B_z^i(x,y)F_H(x,y)}{\int_A {\rm d}x {\rm d}y F_H(x,y)} \,.
\label{eq_2}
\end{equation} 
Here, $\langle B_z(x,y) \rangle$ is the stray field detected in the plane of the 2DEG buried about 115\,nm below the Cr/Au top-gate onto which the magnetic structures are grown. $A$ is the active area of the Hall cross and $B_z^i(x,y)$ is the stray field of the $i$-th nano-cube/-tree in the respective $2 \times 2$-array detected in the 2DEG plane.

In this work, we have assumed a constant $F_H(x,y) \approx 1$ resulting in $\langle B_z(x,y) \rangle \approx 1/A \cdot \sum_{i = 1}^4\int_A {\rm d}x {\rm d}y B_z^i(x,y)$. Although calculating $F_H(x,y)$ requires extensive numerical simulations for each magnetization configuration, we can roughly estimate that the expected decrease of the Hall response may be up to 30\,\% for the present geometry \cite{Peeters2002}. Another large effect is the increase of the effective Hall cross area $A$ due to the rounding of the corners. For both the macro-spin and micromagnetic simulations an idealized quadratic shape have been used with channel widths taken from the SEM microgrpahs, i.~e.\ neglecting round corners and edge depletion effects. Assuming a realistic increase of $A$ by a factor of about 1.3 reduces the calculated stray fields by another 50\,\% for $\theta = 0^\circ$.
Other factors that may contribute to the discrepancy between the measured and calculated stray fields are a small uncertainty in determining the distance to the 2DEG (an effect, however, less than 5\,\%), and possible uncertainties in determining the magnetic volume of the sample and the exact value of the saturation magnetization of the material. Finally, the calculation of the stray field from the measured differential Hall voltages do not account for possible inhomogeneous current distributions in the active area of the Hall cross.   
%
%
\section{Comparison of micromagnetic and macro-spin simulations}
\begin{figure}[htb]
\centering
\includegraphics[width=0.95\columnwidth]{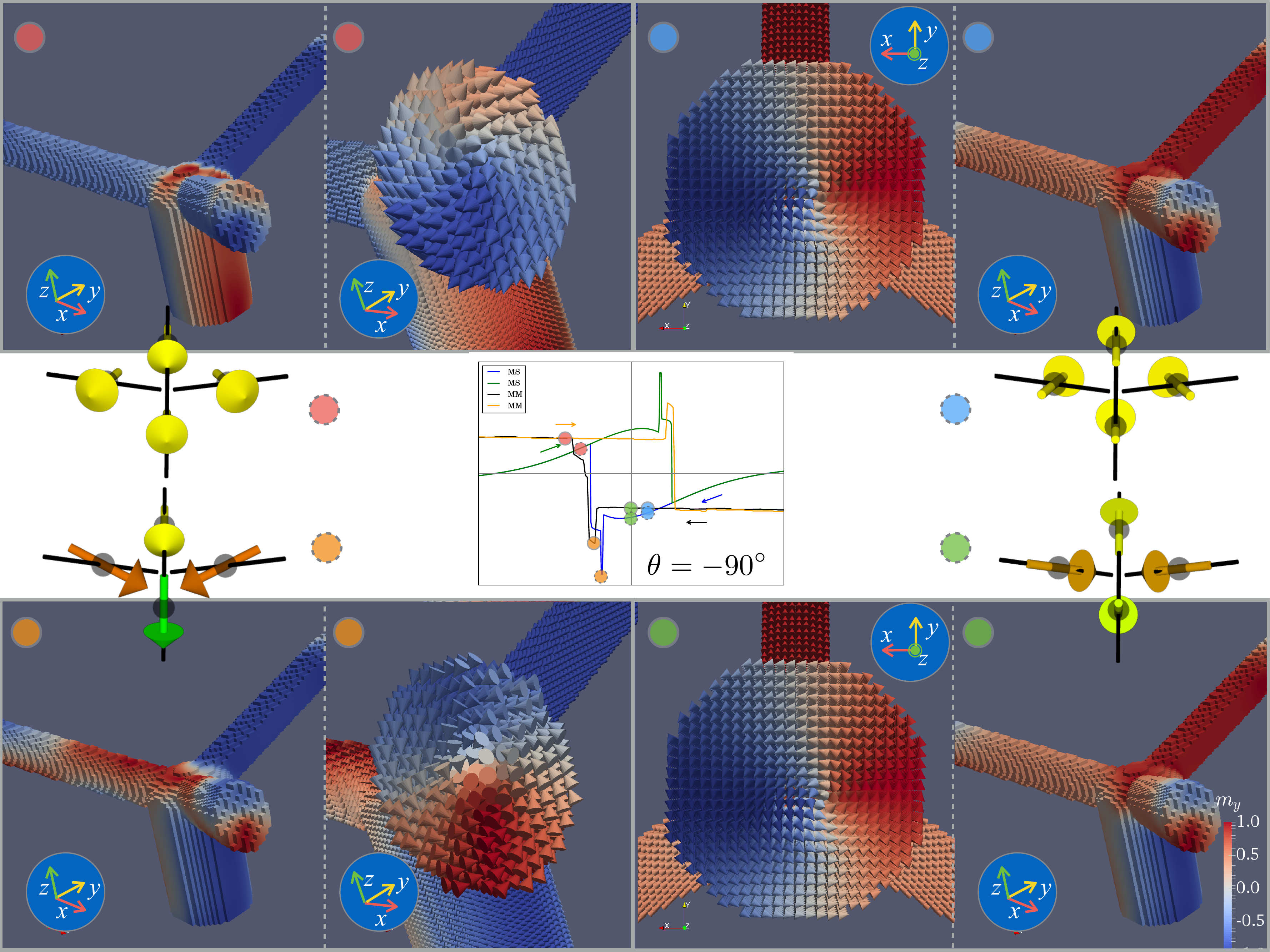} 
\caption{{\bf Comparison of macro-spin and micromagnetic simulations of a pure CoFe nano-tree.} In the center the hysteresis loops for $\theta = -90^\circ$ are shown. Solid- and dashed-line circles mark the positions in the hysteresis loops where micromagnetic and macro-spin configurations, respectively, are shown. Arrows indicate the directions of the field sweeps. The color bar in the lower right sub-plot indicates the color-code for the magnetization's $y$-component (field direction): red -- magnetization fully in field direction, blue -- magnetization fully opposite to field direction. The same color coding is used for the sub-plots that show the respective magnetization direction by cones. For comparison, the macro-spins' orientations at the selected states indicated by the dashed-line circles are also shown. The colors of the macro-spins relate to the colors for the $x$-, $y$- and $z$-axis shown within the blue discs.}
\label{without}
\end{figure}

\begin{figure}[htb]
\centering
\includegraphics[width=0.95\columnwidth]{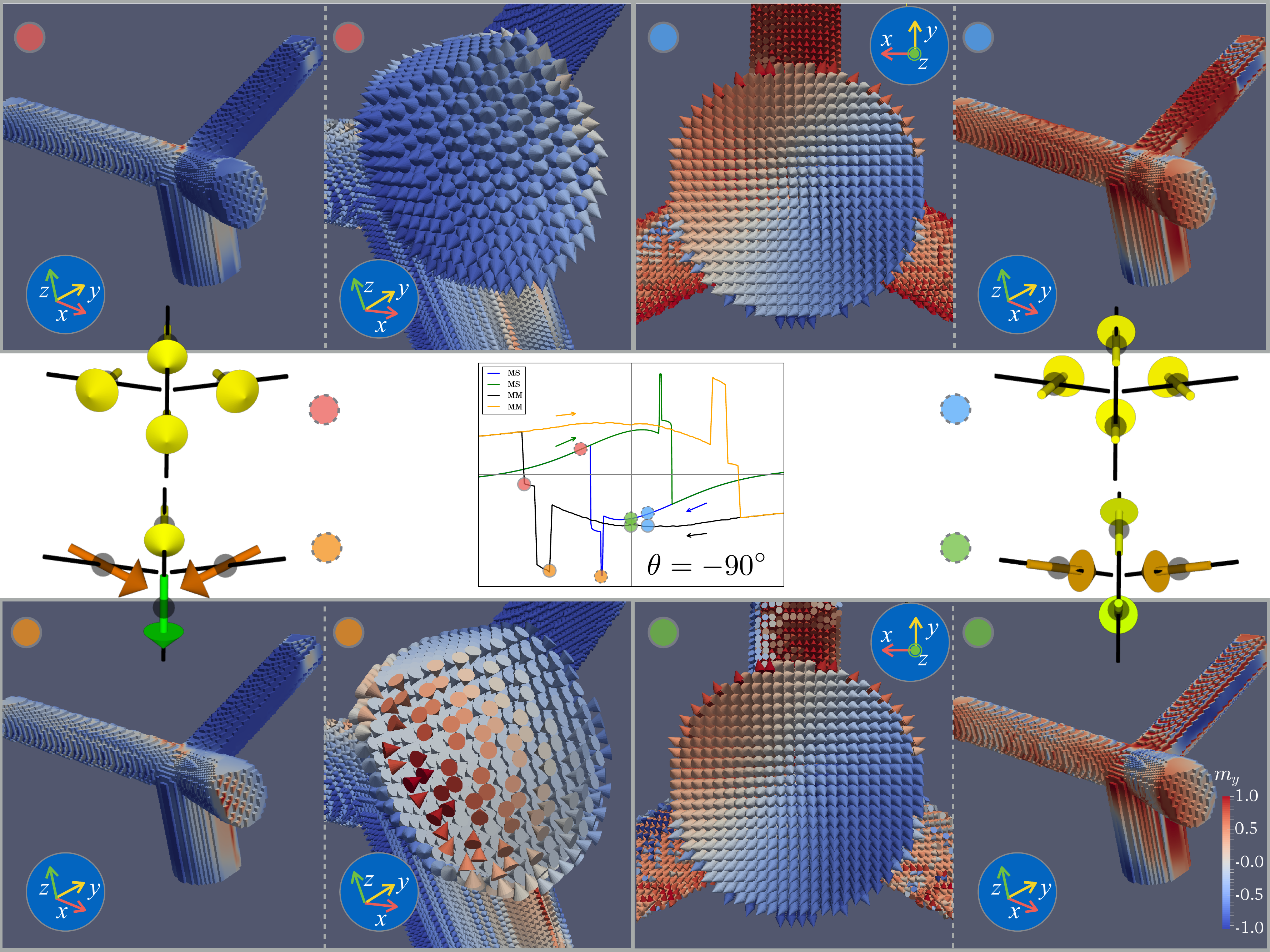} 
\caption{{\bf Comparison of macro-spin and micromagnetic simulations of a CoFe nano-tree with a metal-oxide ferrimagnetic shell.} In the center the hysteresis loops for $\theta = -90^\circ$ are shown. Solid- and dashed-line circles mark the positions in the hysteresis loops where micromagnetic and macro-spin configurations, respectively, are shown. Arrows indicate the directions of the field sweeps. The color bar in the lower right sub-plot indicates the color-code for the magnetization's $y$-component (field direction): red -- magnetization fully in field direction, blue -- magnetization fully opposite to field direction. The same color coding is used for the sub-plots that show the respective magnetization direction by cones.  For comparison, the macro-spins' orientations at the selected states indicated by the dashed-line circles are also shown. The colors of the macro-spins relate to the colors for the $x$-, $y$- and $z$-axis shown within the blue discs.}
\label{with}
\end{figure}

SI-Figures\,\ref{without} and \ref{with} compare the micromagnetic (MM) and macro-spin (MS) simulations for a single CoFe nano-tree with and without the metal-oxide sheath consisting of a ferrimagnetic spinel phase. 

The MS and MM simulations without the core/shell structure, shown in SI-Fig.\,\ref{without}, are very similar. One is therefore led to assume that the macro-spin model contains the essential features of the magnetization reversal process. However, the MM simulations show quite clearly that this assumption is premature. For the prominent stray field states (see solid- and dashed-line circles in the figure) selected, vortex-like magnetization profiles are visible at the terminal faces of the cylinder-shaped edges and also at the bottom of the stem. A closer inspection of different cross section through the nano-tree (not shown) reveals that these vortex structures are not threading throughout the full sample volume on any given cylindrical element. This is caused by the magnetic vertex segment joining the three edges and the stem of the nano-tree. 

The situation is even more complex, if the nano-tree's core-shell structure is taken into account. In this case, the distribution of magnetization orientations seen in the MM simulations exhibits even stronger spatial inhomogeneities in the magnetization direction. At the end caps of the cylindrical edges hedgehog-like structures occur, whereas the vortex-like magnetization profiles observed for the all-metal nano-tree micromagnetic model are virtually absent. We note that the thickness of the spinel outer layer only contains two voxel cells. A fully satisfying account on the details of the magnetization distribution in the shell would require a voxel edge length significantly below the used 5\,nm. However, even at a moderate reduction to 3\,nm the simulations could not be performed anymore on the hardware available to us (see method section of main text).
%
%
\section{Comparison of magnetic stray field measurements and micromagnetic simulations for 3D nano-cubes}
\begin{figure}[htb]
\centering
\includegraphics[width=0.95\columnwidth]{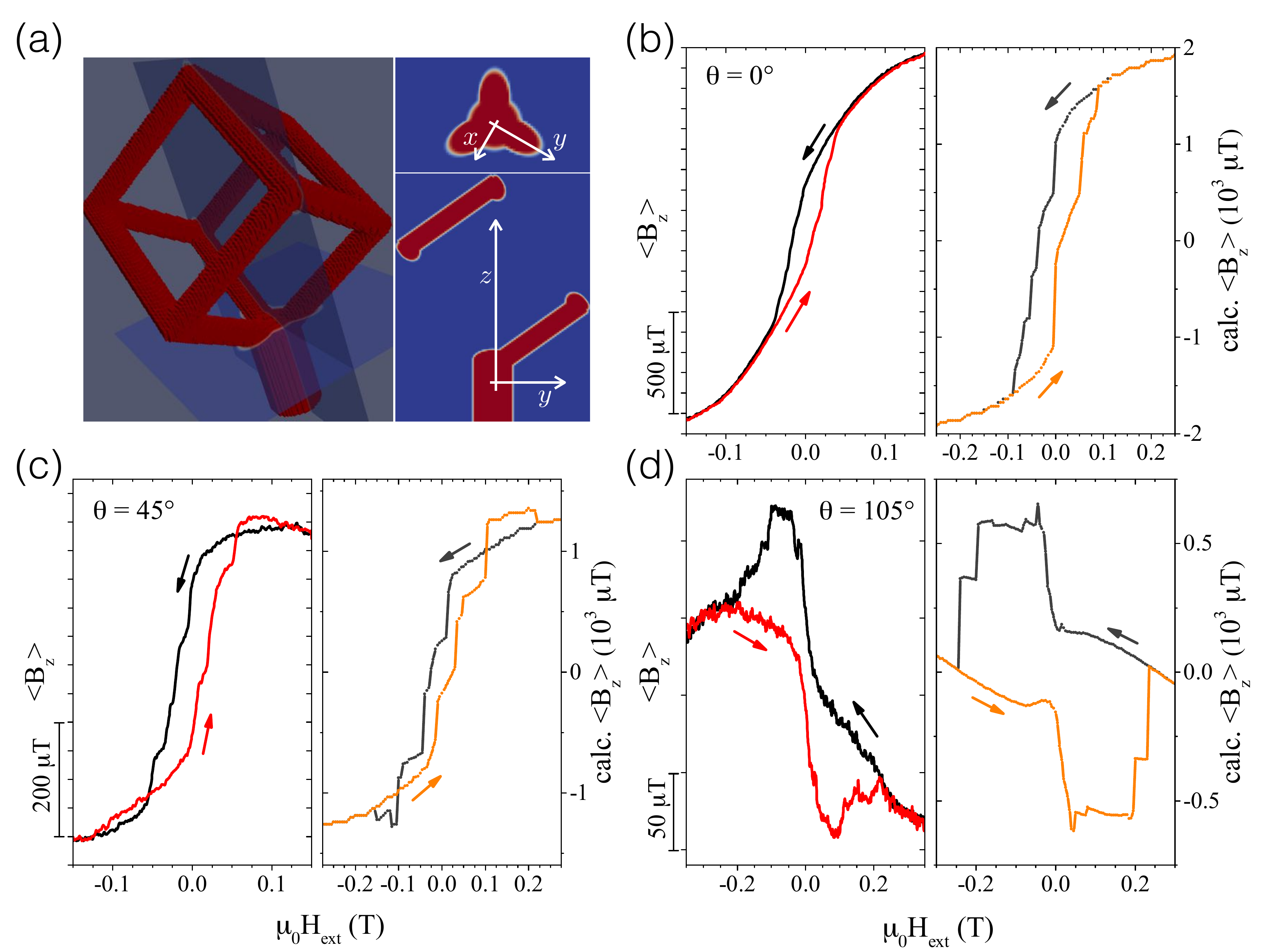} 
\caption{{\bf Comparison of stray field measurements and micromagnetic simulations of a all-metal Co$_3$Fe nano-cube} (a) 3D-view and cross sections of nano-cube assuming a fully metallic Co$_3$Fe volume. The cross section planes are indicated in the 3D view. (b), (c) and (d) Comparison of the measured magnetic stray fields $\langle B_z \rangle$ at $T = 30$\,K and micromagnetic simulations for $T = 0$\,K for selected angles $0^\circ$, $45^\circ$ and $105^\circ$, respectively. Arrows indicate directions of field sweeps.}
\label{fig_SI_cubes_vgl_MM}
\end{figure}

In SI-Fig.\,\ref{fig_SI_cubes_vgl_MM} we show the results of micromagnetic simulations at the same angles which are shown in Fig.\,4 of the main paper. The all-metal micromagnetic model already shows a very good qualitative agreement with the measured stray fields. For example, the experimentally observed crossing of the up- and down-sweep curves for $\theta = 45^\circ$ are well reproduced by the simulations. As we have shown for the nano-trees, taking into account a core/shell structure with an outer metal-oxide layer we would expect an even better agreement with the measurements. However, for the nano-cube micromagnetic model the simulation volume is significantly larger than for the nano-tree. Even at the rather large 5\,nm edge length for the cubic voxels used in our simulations, a core/shell structure simulation could not be performed anymore on the hardware available to us (see method section of main text).
%
%
\end{document}